\documentclass[journal=jacs,manuscript=article,layout=twocolumn]{achemso}
\usepackage[version=3]{mhchem} 
\usepackage{amsmath,amssymb}
\usepackage{bm}
\usepackage{multicol}

\let\oldmaketitle\maketitle
\let\maketitle\relax

\author{Shensheng Chen}
\email{shensheng@caltech.edu}
\affiliation{Division of Chemistry and Chemical Engineering, Caltech}
\author{Tingtao Zhou}
\email{edmondztt@gmail.com}
\affiliation{Division of Engineering and Applied Science, Caltech}

\title{Polyelectrolyte Knot Delocalization Induced by Counterion Condensation}

\begin{document}


\begin{tocentry}
\vfill
  \begin{minipage}{1\textwidth} \end{minipage}%
  \raisebox{-.1\dimexpr\height-\ht\strutbox+\dp\strutbox\relax}%
  {\includegraphics[width=1\textwidth]{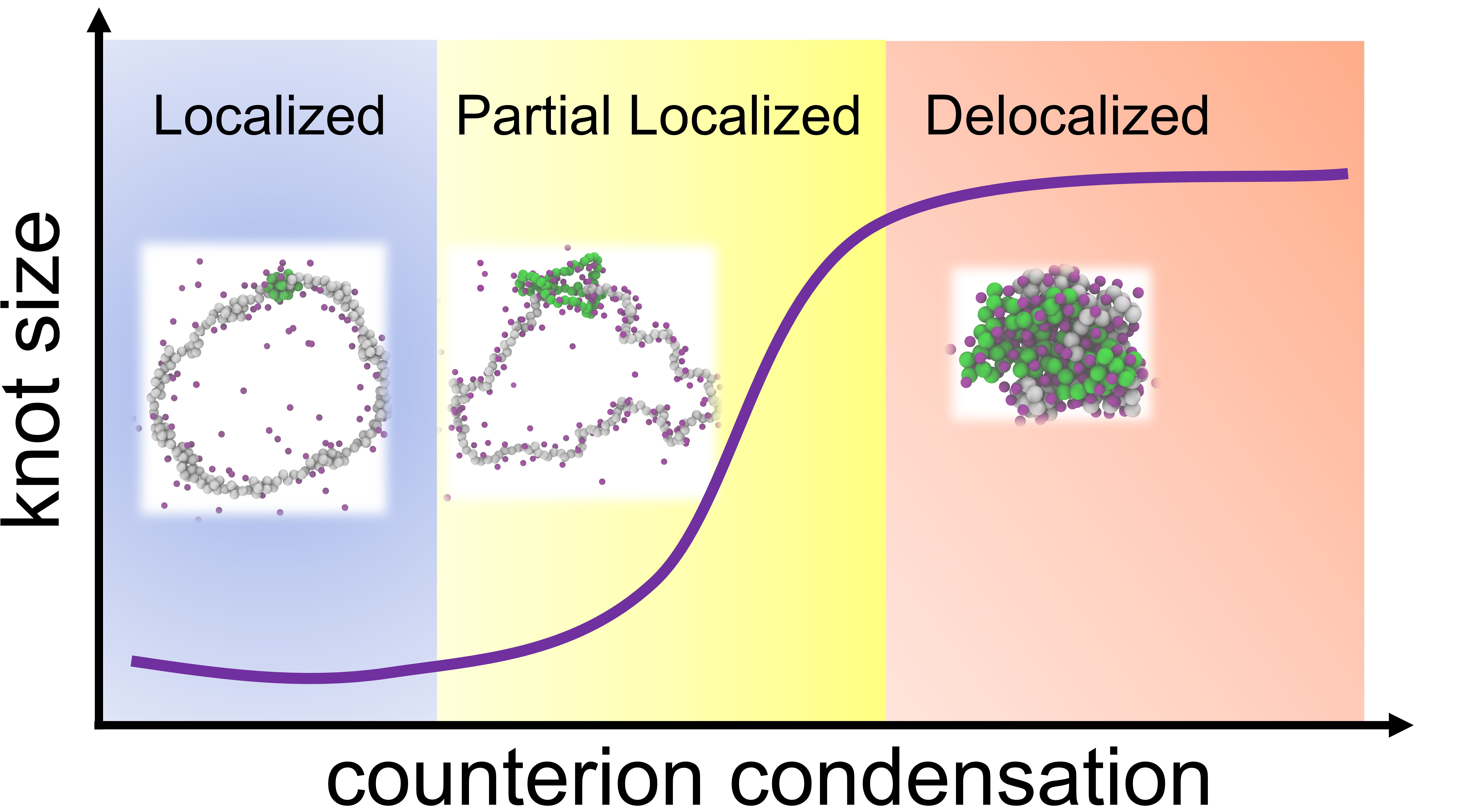}}
\\\vfill
\end{tocentry}

\twocolumn[
\begin{@twocolumnfalse}
\oldmaketitle
\begin{abstract}
Knots on ring polymers tend to be tight. For knots on charged polymers (polyelectrolytes), electrostatic repulsion among the monomers is considered to enhance the persistence length and further localize the knots. However, the effects of counterion condensation on knots behavior are not clear. Here we use molecular dynamics simulations to systematically study the effect of counterion condensation on the knot behavior under different electrostatic strength and solvent quality conditions, with the focus on the knot sizes. We show that generally counterion condensation \emph{delocalizes} the knots in systems with strong electrostatic strength. At small to intermediate electrostatic strength, the knot size and overall chain size and can be positive correlated or negative correlated, as a result of the interplay between electrostatic correlation and solvent quality.

\end{abstract}
\end{@twocolumnfalse}
]

Polymer knot is an important yet elusive aspect of topological interactions in polymer physics~\cite{wang201750th,micheletti2011polymers,kardar2008elusiveness,dai2014metastable,dai2020developing}. Probabilistic arguments and numerical simulations have shown that knots should occur easily for sufficiently long chain length~\cite{sumners1988knots,van1990knot,tesi1994knot,shimamura2002knot} or densely packed configurations~\cite{mansfield1994knots,virnau2005knots,baiesi2011topological,amin2018nanofluidic}. In biology, knots have been reported for DNA plasmid, viral DNA genomes and proteins~\cite{meluzzi2010biophysics}. Artificial nanoscale knots have been also fabricated in recent years for potential applications in molecular machines~\cite{dietrich1989synthetic,arai1999tying,bao2003behavior,ayme2012synthetic,beves2011strategies}.

A key problem is the size of these knots on the polymers, since it can affect the morphology, mechanical/rheological properties and potentially biological functionality of proteins~\cite{dabrowski2016search}. Knots on a Gaussian polymer are usually considered to be tight due to the entropy maximization, under the slip-link argument~\cite{duplantier1986polymer,duplantier1989statistical,metzler2002tightness}. Monomer interactions have been shown to affect the knot size in linear chains~\cite{dai2016effects}. For knotted polyelectrolytes (charged polymers) such as most bio-molecules,  previous studies suggested that the bare coulomb interaction  between monomers always tightens the knots.~\cite{dommersnes2002knots} and this electrostatic-induced knot localization effect persists with added salt as long as the electrostatic persistent length $\ell_p = \ell_B (\lambda/b)^2$ is larger than the polymer size $N$~\cite{kardar2008elusiveness}. Here, $\ell_B$ is the Bjerrum length and $\lambda$ is the screening length that decays with salt concentration. Based on mean-field augments, these  prediction highlights that electrostatic interaction leads to knot tightening. However, the effects of electrostatic correlation such as counterion condensation on the knots behavior is not clear.

For a knot-free polyelectrolyte, counterion condensation takes place when the electrostatic strength $\ell_B/b >1$ with $b$ being the charge spacing of two neighbored monomers along the chain~\cite{muthukumar2004theory,liao2003molecular,dobrynin2005theory,deshkovski2001counterion,liao2006counterion}. Counterion condensation increases with Bjerrum length, and will eventually lead to the coil-to-globule transition of the polyelectrolyte~\cite{muthukumar2004theory,liu2002langevin,kundagrami2010effective,muthukumar201750th}. In aqueous environments, some bio-molecules such as DNA have high charge density $\ell_B/b > 3$, where the counterion condensation is expected to be significant. The study of electrostatic correlation on knot behavior is nearly untouched. Recently, \citep{tagliabue2020interface} has shown the morphology (quantified by the radius of gyration of the whole chain) of a trefoil-knotted polyelectrolyte is significantly influenced by the condensed counterions at high $\ell_B$. This study highlights the change in the radius of gyration of the knotted chain. However, the role of counterion concentration on the local knot behavior, especially the influence on the knot size, is not clear.

In this work, using Molecular Dynamics simulations, we study the effect of counterion condensation on the local knot size of a polyelectrolyte ring under different electrostatic strength and solvent conditions. The short-range interaction between two beads in our simulations is based on the Kremer-Grest (KG) model~\cite{grest1986molecular,kremer1990dynamics} with the non-bonded interaction given by  $U_{LJ}^{ij}(\bm{r_{ij}})=4\varepsilon_{ij}\left((\frac{b_{ij}}{{r_{ij}}})^{12}-(\frac{b_{ij}}{{r_{ij}}})^{6}\right)$, with a cut-off distance $r_{cut}=2.5~b_{ij}$ for interaction between polymer beads and $r_{cut}=2^{1/6}~b$ for others. We set $b_{ij}=b$ for all beads. To control the solvent quality, we vary $\varepsilon_{ij}= 0.5~k_BT \sim 3.5~k_BT$, where $k_BT$ represents the thermal energy and serves as the energy unit in the systems. The corresponding solvent-polymer Flory-Huggins parameter reads $\chi=0.5*(1-B_2/v)=1.2372 \sim 57.46$, where $B_2$ is the second virial coefficient given by $B_2=-2\pi\int (e^{-U_{LJ}(r)/k_BT}-1)r^2dr$ and $v=\frac{4\pi}{3}(b/2)^3$ is the volume of a monomer. The bonded interaction is described by the FENE potential $U_{FENE}^{ij}(r_{ij})=\frac{1}{2}K_{bond}R_{0}^2[1-(\frac{r_{ij}}{R_0})^2]$, with $K_{bond}=30~k_BT/b^2$ and $R_0=1.5~b$. We simulate a trefoil-knot ring with $N=200$ monomers and $200$ counterions. Each monomer carries unit charge $-e$, and each counterion carries the unit charge $e$ to maintain the overall charge neutrality. The long range electrostatic interaction is given by $E_{el}^{ij}= k_BT \frac{l_Bq_{i}q_{j}}{ r_{ij}}$, where $q_\alpha$ is the charge on bead $\alpha$. In this study, the effect of electrostatic strength is investigated by varying Bjerrum length $\ell_B=1~b \sim 10~b$. The length scale $b$ and energy scale $k_BT$ are both set to unity in simulation. The simulation time scale is given by $\tau=\sqrt{mb^{2}/k_{B}T}=1$. We first equilibrate each simulation for $10^5 \tau$ and perform data collection for additional run of  $10^6 \tau$. 

To identify the location and size of the knot, we adopt the knot search method by Tubiana et al.~\cite{tubiana2011probing,tubiana2018kymoknot} : the ring configuration is cut open and shrinks from the two ends as the search proceeds, and then connected to form a new ring via the minimally-interfering method~\cite{tubiana2011probing}. The knot size is defined as the minimum number of consecutive monomers that preserves the trefoil topology, denoted by $N_k$.
As reference for later discussions, the minimum size of an `ideal' closed trefoil knot is about 16.372~$b$~\cite{stasiak1998ideal,denne2006quadrisecants,sullivan2002approximating,rawdon2003can}.



\begin{figure}[tbh]
\centering
\includegraphics[width=1\columnwidth]{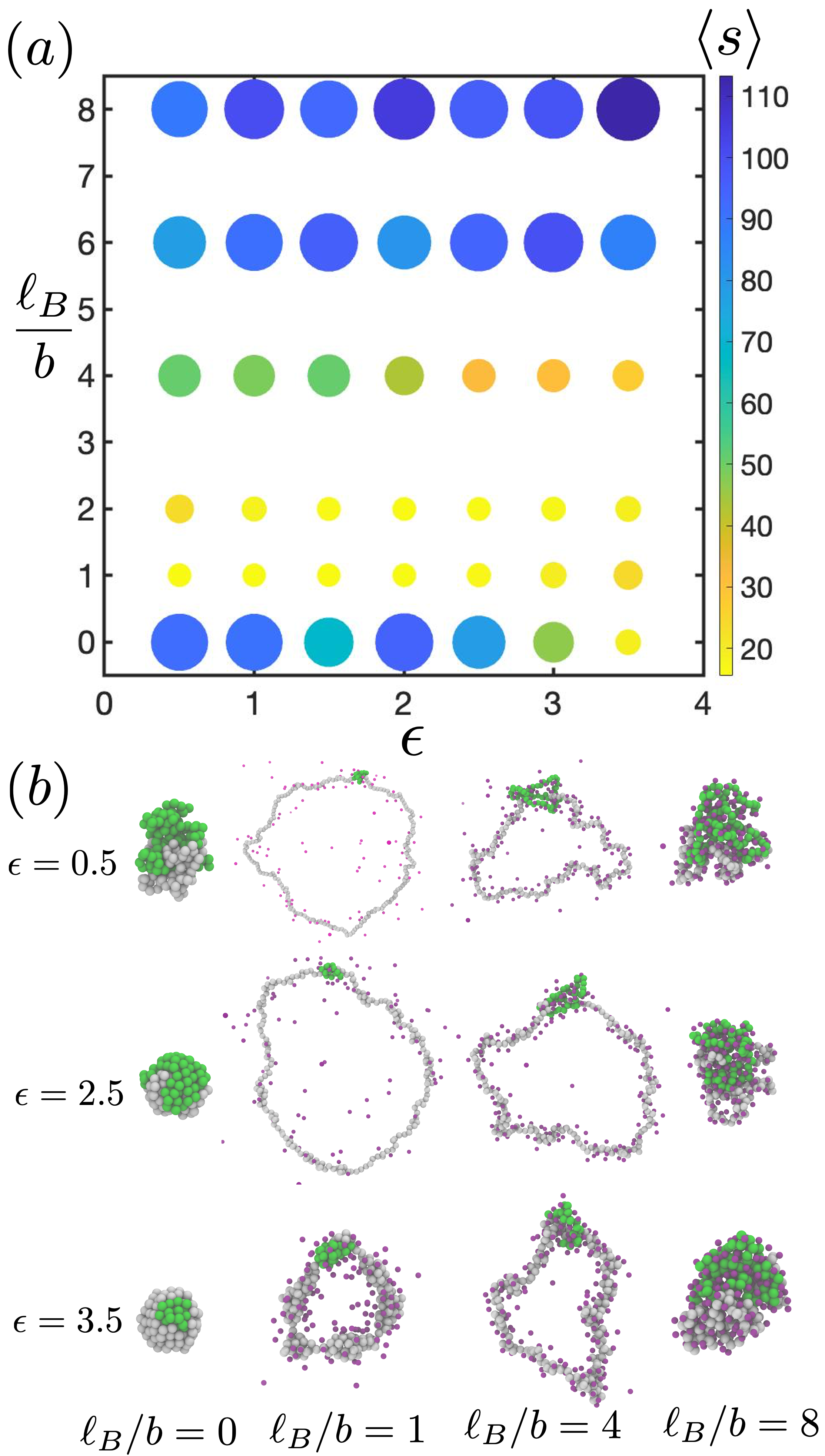}
\centering
\caption{\label{fig:knot-szie phase diagram}
(a) Phase diagram for the knot size as the electrostatic strength $\ell_B/b=e^2/(4\pi\varepsilon k_BT b)$ and solvent quality $\epsilon$ varies. Both colors and sizes of the circle correspond to the knot sizes. For the charged knots,
a phase boundary shows up around $\ell_B/b=5$. For $\ell_B/b \lesssim5$ in solvent quality $\epsilon <2$, partial counterion condensation takes place and the knot can be as large as three times compared to the tightest configuration. 
(b) Typical knot configurations.
The knotted portion of the ring polymer is highlighted in green, with the rest of the monomers in grey, and counterions are colored dark pink. The neutral knots are shown here ($\ell_B=0$) for reference, which under poor solvent conditions are crumpled globules with large knot sizes. }
\end{figure}

\begin{figure}[h]
\centering
\includegraphics[width=0.9\columnwidth]{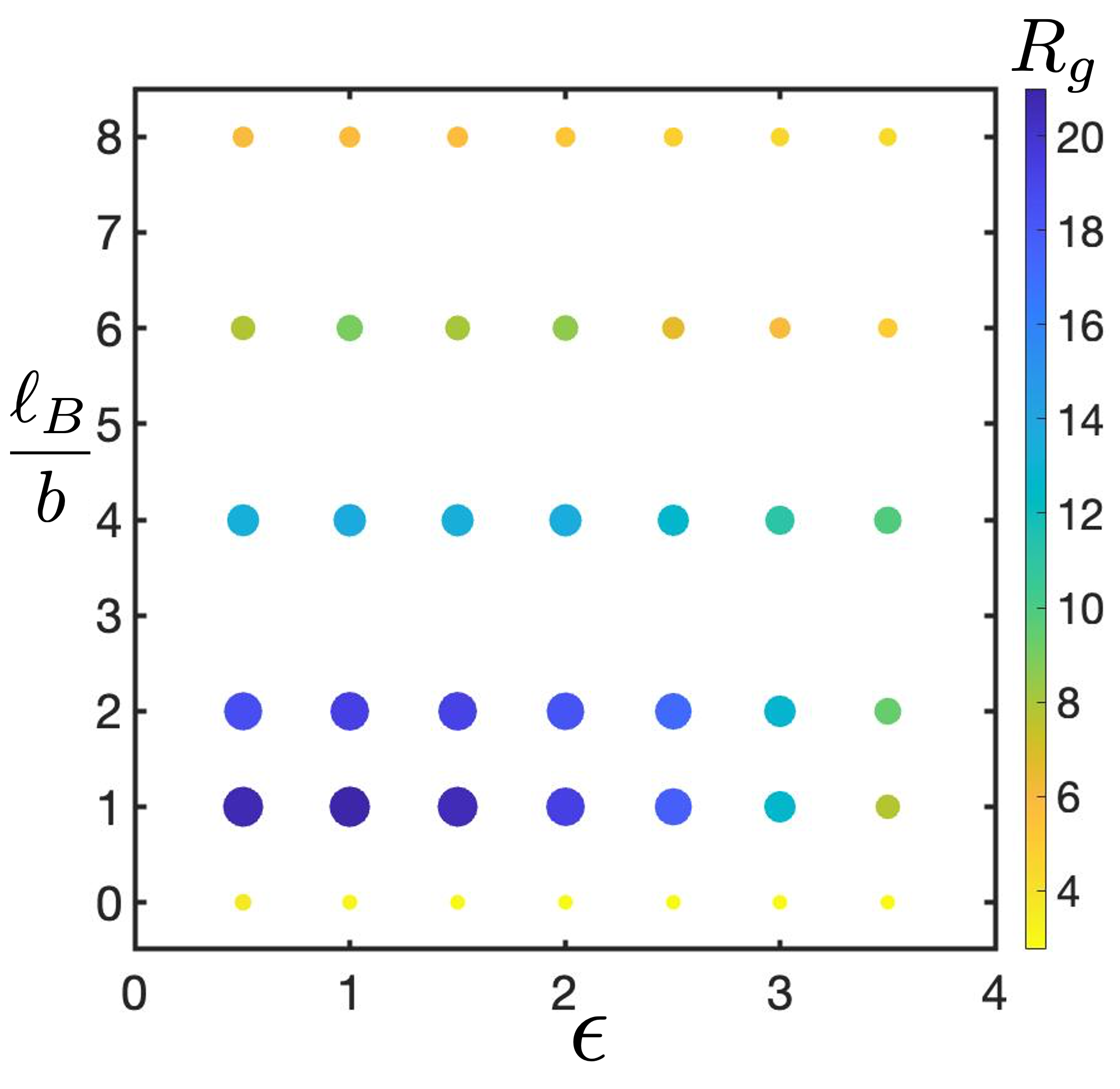}
\centering
\caption{\label{fig:Rg}
The radius of gyration $R_g$ as $\ell_B/b$ and $\epsilon$ varies. For large $\ell_B/b$, delocalized knots are accompanied by dense crumpled globule states, with $R_g$ smaller than that of a free Gaussian ring. For small $\ell_B/b$, poor solvent quality significantly reduces the $R_g$, enhancing counterion condensation and promote denser packings as shown in Fig.~\ref{fig:knot-szie phase diagram}(b) and Fig.~\ref{fig:gofr}. Notice that for very small $\ell_B/b=1$, $R_g$ first decreases then increases as $\epsilon$ increases, reflecting the morphology of a elongated crumpled rod at very poor solvent quality.
}
\end{figure}

\begin{figure}[h]
\centering
\includegraphics[width=0.9\columnwidth]{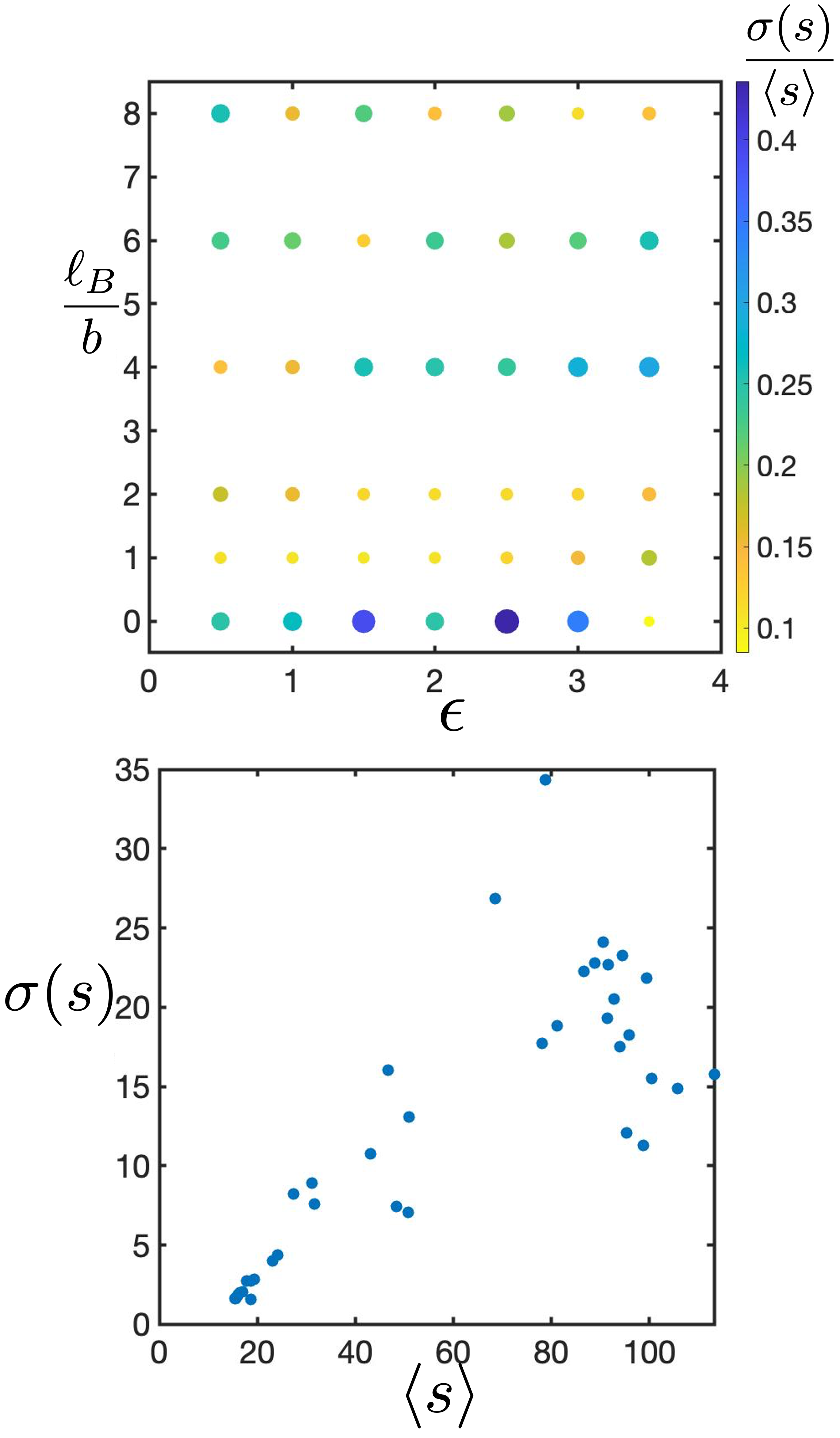}
\centering
\caption{\label{fig:flucutations}
Knot sizes exhibit giant fluctuations in the delocalized regime, indicating that delocalized knots are a very dynamic phase. $\left<s\right>$ is the time averaged knot size, $\sigma(s)$ is the standard deviation.
}
\end{figure}

We first demonstrate a localization-delocalization phase boundary determined solely by $\ell_B/b$ in Fig.~\ref{fig:knot-szie phase diagram}, the corresponding phase diagram of radius of gyration of the whole chain is given at Fig.~\ref{fig:Rg} (a). As shown in Fig.~\ref{fig:knot-szie phase diagram}(a), all the knots in configurations with $\ell_B/b >5$ are delocalized, with sizes about half of the total chain length. This knot delocalization phenomenon at large Bjerrum length can not be explained by mean-field argument that predicts knot tightening due to electrostatics, rather it is a consequence of counterion condensation due to strong electrostatic correlation. Indeed, as visually shown in Fig.~\ref{fig:knot-szie phase diagram}(b), the counterions are almost entirely condensed at $\ell_B/b=8$. The dipole--dipole fluctuation interaction results in a polyelectrolyte globule, and consequently, a relaxed knot. In the regime of ( $\ell_B/b>5$), the whole ring essentially clumps into a globule due to strong electrostatic correlation, as confirmed by the nearly constant radius of gyration $R_g$ of the whole chain shown in Fig.~\ref{fig:Rg}. The knotted portion and non-knotted portion of beads clump together, yet they are not fully mixed due to topological constrain of the knot, leading to clearly regional separation. 
In this regime, while the knot size stay almost the same, the overall radius of gyration decreases with worsen solvent quality as a result of compacting effect.
We remark that the delocalized knots are a very dynamic phase, manifested by the giant fluctuations of the knot sizes, as show in Fig.~\ref{fig:flucutations}.

\begin{figure*}[tbh]
\centering
\includegraphics[width=1\textwidth]{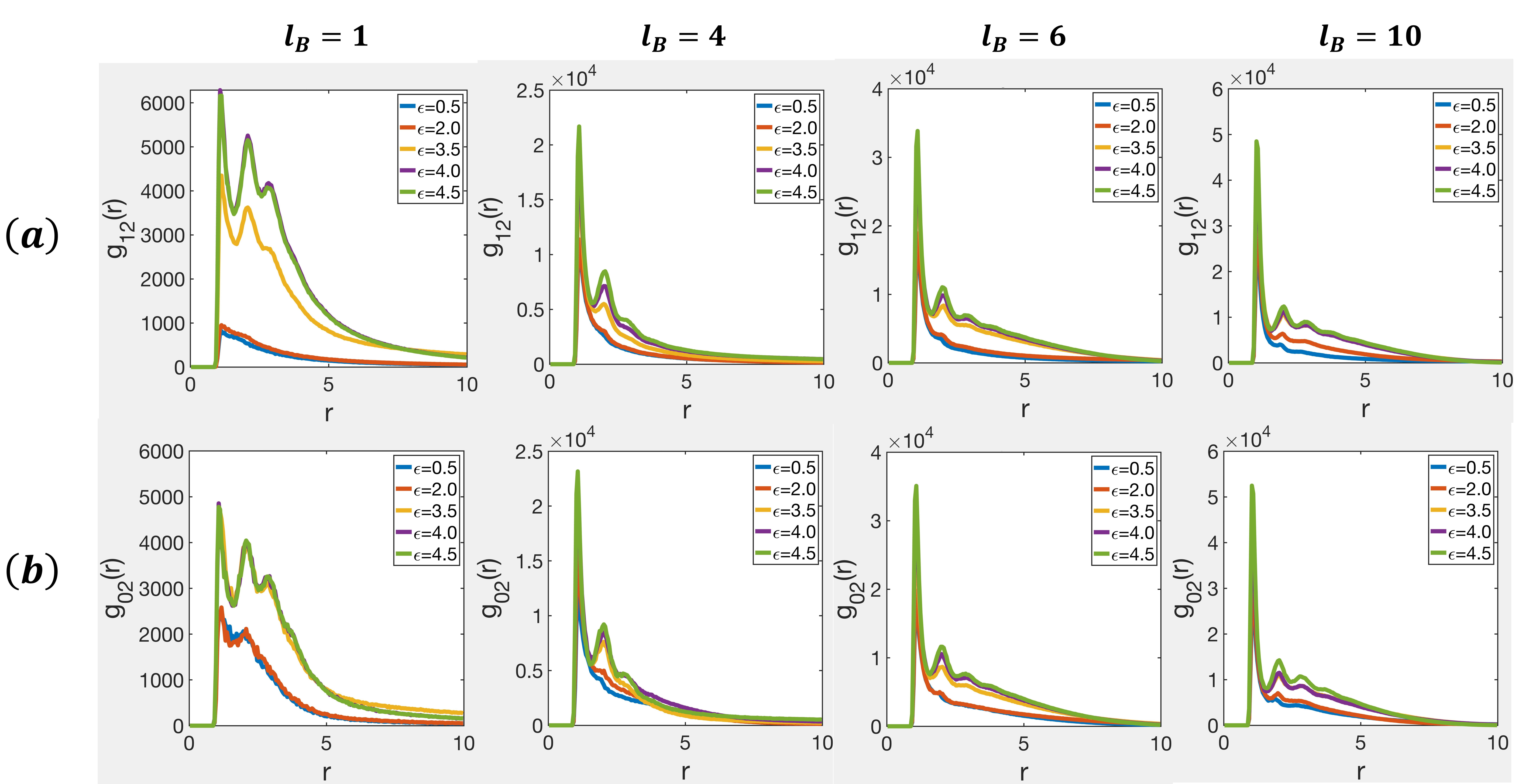}
\centering
\caption{\label{fig:gofr}
Counterion distributions characterized by the 
radial distribution function g(r) for pairs of knotted monomer-counterions (row (a) g$_{12}$) and unknotted monomer-counterions (row (b) g$_{02}$). 
}
\end{figure*}

For $\ell_B/b<4$, the knots are mostly tight, In this regime, the degree of ionization is still high (will discuss more later in Fig.~\ref{fig:gofr}), the net charge along the polymer backbone still tightens the knot. In this regime, the knot size slighly increases with worsen solvent quality, due to the fact that poorer solvent induce effective attraction between monomers that counteracts the electrostatic repulsion. As a result, worsen solvent quality induce deceasing of the overall ring size $R_g$ (as shown in ~\ref{fig:Rg} (a)), but increases the knot size due to slight knot delocalization. As shown in Fig ~\ref{fig:Rg} (b), in this regime ($\ell_B/b=1$ and $\ell_B/b=2$), the knot size and $R_g$ are negative correlated, as expected. Interestingly, at $\epsilon=1$, the first row of Fig.~\ref{fig:knot-szie phase diagram}(b) shows that the ring maintains its shape in low $\epsilon$ (better solvent quality). As solvent quality decreases, it transits into a crumpled necklace state~\cite{kantor1994excess}, and eventually collapses into a crumpled rod with much more condensation of counterions at very bad solvent condition ($\epsilon=4.5$). These morphology features are consistent with previous predictions~\cite{kantor1994excess,khokhlov1980collapse,raphael1990annealed,dobrynin1996cascade}.      

Interestingly, at the intermediate electrostatic strength $\ell_B/b=4$ where the counterions are partially charged, the knots clearly decreases with worsen solvent quality, which is different from the case in $\ell_B/b=1$ and $\ell_B/b=2$. As shown at Fig.~\ref{fig:Rg} (b), at $\ell_B/b=4$, the radius of gyration of the ring deceases with worsen solvent quality, follow the same trend as the case in $\ell_B/b=1$ and $\ell_B/b=2$. Thus, in this intermediate regime, the knot size and $R_g$ are positive correlated, as shown in Fig ~\ref{fig:Rg} (b). This is a result of the intrigue effect of the interplay between electrostatic correlation and solvent quality on the local knot behavior: In this regime at small $\epsilon$, counterion condensation partially delocalize the knots (Fig.~\ref{fig:knot-szie phase diagram} (a)), resulting in a slighly loose knot on the ring. When solvent quality deceases, its induced attraction among all monomers effectively scale the ring with a smaller diameter and a thicker tube on the perimeter. As a result, the neighbor non-knotted beads on the two sides of the knot attracts the knot beads and consequently `squeezes' the knot.

To further understand the interplay between electrostatic strength and solvent quality on the knot behavior, we analyze the ion distribution, as shown in Fig.~\ref{fig:gofr}. In Fig.~\ref{fig:gofr}, we use g$_{12}$ to represent the pair correlations between knotted monomers and counterions (row (a)), and g$_{02}$ to represent unknotted monomers and counterions (row (b)). The first peak magnitude of g(r) indicates the amount of condensed counterions. 

At small $\ell_B/b=1$ with low $\epsilon$, the degree of counterion condensation is low, as seen in the low first peaks of both g(r), and g$_{12}$ resembles a dilute gaseous phase. The difference between g$_{12}$ and g$_{02}$ is due to the very tight knot, which creates a significant excluded volume for the counterions. As solvent quality further decreases, the degree of counterion condensation increases, and eventually the knotted monomers shows a higher first peak in g(r). These can be understood by the two competing effects of surface tension and electrostatics: the poor solvent quality try to squeeze the whole ring while the electrostatic repulsion between the monomers try to swells the ring. The worse solvent quality shrinks the ring and consequently significantly increases the electrostatic penalty that enhances counterion condensation. The original tight knot area attracts more counterion to maximize conformation entropy.

For delocalized knots at $\ell_B/b>5$, due to the significant counterion condensation induced by strong electrostatic correlation, the polyelectrolyte ring is always in the crumpled globule state, and there is no difference between knotted and unknotted beads locally, as shown by the similar  g$_{02}$ and g$_{12}$. The degree of counterion condensation increases continuously with $\ell_B/b$, and is not much affected by solvent quality change, evidenced by the overlap of the first peaks. Very bad solvent always squeezes the system into a densely packed liquid state, showing significant second and third peaks in both g$_{12}$ and g$_{02}$ for all cases, and visible higher-order peaks for $\ell_B/b>5$. 

The radius of gyration shown in Fig.~\ref{fig:Rg} (a) is overall negatively correlated with the knot size: intuitively, $R_g$ is small when the polyelectrolyte is crumpled with shorter persistence length, and the knot is relaxed. For $\ell_B/b \le4$ in very poor solvent, $R_g$ significantly decreases as the crumpled necklace or cylinder state emerges, while the knot still remains tight. Thus, overall, solvent quality has more impact on the overall morphology of the ring than the knot size.

In summary, we have performed Molecular Dynamics simulations to demonstrate the effect of counterion condensation on the knot size and morphology of a trefoil-knot ring polymer under different electrostatic strength and solvent quality conditions. We highlight that strong electrostatic correlations leads to counterion condensation, which delocalizes the knot. The knot size of the overall chain size can be positive or negative correlated. This result complements the mean-field theory of knot localization. Demonstrated on a trefoil knot, we suggest this condensation-delocalization mechanism to be generally true for more complicated knots, either prime or composite ones. Future works are needed to understand  effects of various factors on knot behaviors in polyelectroltyes, such as the added salt, chain length, and charge density and charge distributions along the chain.   
We notice that during the review process of this manuscript a similar work is published~\cite{tagliabue2022tuning}.

\begin{acknowledgement}

The authors thank Z.-G. Wang for discussions. T. Z. acknowledges support of the Drinkward Postdoc Fellowship during part of this work.

\end{acknowledgement}




\bibliography{refs}

\end{document}